\documentclass{article}
\usepackage[utf8]{inputenc}
\usepackage{geometry}
\usepackage{graphicx}
\usepackage{amsmath, amssymb}
\usepackage{mathtools}
\usepackage[sorting = none]{biblatex}
\usepackage{booktabs}
\usepackage[colorlinks=true, linkcolor=blue]{hyperref}

\title{Finding the Maximal Contrast of Two Elliptical Gaussian Mode Beams with Aligned Ellipticities}
\author{Mark Zhu, Sina M. Koehlenbeck, Edgard Bonilla, Brian Lantz}
\date{June 2025}

\begin{document}

\setlength{\marginparwidth}{0pt}
\setlength{\marginparsep}{5pt}

\maketitle
\begin{abstract}
    Interferometric contrast is a key factor limiting the sensitivity of precision optical measurements, including the laser interferometers used in gravitational-wave detection. While standard formulas describe the interference of circular Gaussian beams, many real systems use beams with elliptical cross sections, where differing waists and radii of curvature can reduce fringe visibility. This paper derives an analytic expression for the maximum contrast achievable between two aligned elliptical Gaussian beams, written entirely in terms of their geometric and power parameters. We then test the formula using a free-space Michelson interferometer in which all beam parameters are independently measured through beam profiling and nonlinear fitting. In our experiment, the predicted maximum contrast was 0.968 while the experimentally optimized value was 0.950. The small discrepancy is consistent with expected imperfections such as beam rotation, mode mismatch, and non-Gaussian aberrations. This work provides a practical tool for modeling and optimizing elliptical-beam interferometers.
\end{abstract}

\section{Introduction}

The objective of this document is to derive an expression for the `contrast' or `interferometric visibility' of two interfering Gaussian beams with elliptical cross sections which are aligned along the same axis of propagation. This would allow one to calculate a theoretical upper limit to the contrast of an interferometer utilizing elliptical Gaussian beams, which is useful for maximizing the sensitivity and precision of interferometric experiments.\newline

The electric field of a Gaussian beam with a circular cross section can be written as a function of the distance $z$ along its axis of propagation ($\hat{z}$ in this case), the radial distance $r$ away from said axis, and two beam parameters $w_0$ and $z_0$ known as the beam waist and focal position, respectively \cite{Siegman1986}:
\begin{equation}\label{eq:circle}
    \vec{E}_{\text{circ}}(r,z) = E_0\hat{x} \frac{w_0}{w(z)}\exp\left(-\frac{r^2}{w(z)^2}\right)\exp\left[-i\left(kz - \frac{kr^2}{2R(z)} - \psi(z)\right)\right]
\end{equation}

\begin{align}
    \text{Rayleigh Range:} \quad &z_R = \frac{\pi w_0^2}{\lambda} \label{def}\\
    \text{Beam Width:} \quad &w(z) = w_0\sqrt{1+\left(\frac{z-z_0}{z_R}\right)^2} \nonumber \\
    \text{Radius of Curvature:} \quad &R(z) = (z-z_0)\left[1+ \left(\frac{z_R}{z-z_0}\right)^2\right] \nonumber \\
    \text{Gouy Phase:} \quad &\psi(z) = \arctan\left(\frac{z-z_0}{z_R}\right) \nonumber
\end{align}

To describe the electric field of a Gaussian beam with an \textit{elliptical} cross section instead, we need to separately describe the two transverse coordinates $x$ and $y$ rather than grouping them together as $r$. Then each beam is characterized by two beam parameters $w_{0_x}$ and $w_{0_y}$ and three coordinates like so \cite{Arnaud1969}:

\begin{align}\label{eq:ellip}
    \vec{E}(x,y,z) &= E_0\hat{x} \sqrt{\frac{w_{0_x}w_{0_y}}{w_x(z)w_y(z)}} 
    \exp\left(-\frac{x^2}{w_x(z)^2}\right)
    \exp\left(-\frac{y^2}{w_y(z)^2}\right)\notag \\
    &\quad \times \exp\left[-i\left(kz - \frac{kx^2}{2R_x(z)} - \frac{ky^2}{2R_y(z)} - \frac{\psi_x(z)}{2} - \frac{\psi_y(z)}{2}\right)\right]
\end{align}

Note: One can confirm that if $w_{0_x} = w_{0_y}$ then~\eqref{eq:ellip} reduces to~\eqref{eq:circle}\newline

We can define two real variables $A_{xy}$ and $\phi$ which describe the amplitude and phase of an elliptical Gaussian beam:

\begin{align*}
    A_{xy} &= E_0 \sqrt{\frac{w_{0_x}w_{0_y}}{w_x(z)w_y(z)}}
    \exp\left(-\frac{x^2}{w_x(z)^2}\right)
    \exp\left(-\frac{y^2}{w_y(z)^2}\right) \\
    \phi &= kz - \frac{kx^2}{2R_x(z)} - \frac{ky^2}{2R_y(z)} - \frac{\psi_x(z)}{2} - \frac{\psi_y(z)}{2} \\ \\ 
    &\implies \quad E(x,y,z) = A_{xy} e^{-i\phi}
\end{align*}

In this form, we clearly see that the real part of field $E$ is a cosine function. If we interfere two such beams, their fields are superimposed. Since two cosines always sum to another cosine function, the total electric field is also a cosine. Then the total observed power of the interfering beams can be expressed as a cosine function \cite{Wanner2014}:

\begin{equation}\label{contrast}
    P_{\text{tot}} = P_0(1+C\cdot \cos(\phi_{\text{tot}}))
\end{equation}

\begin{figure}
    \centering
    \includegraphics[width = \textwidth]{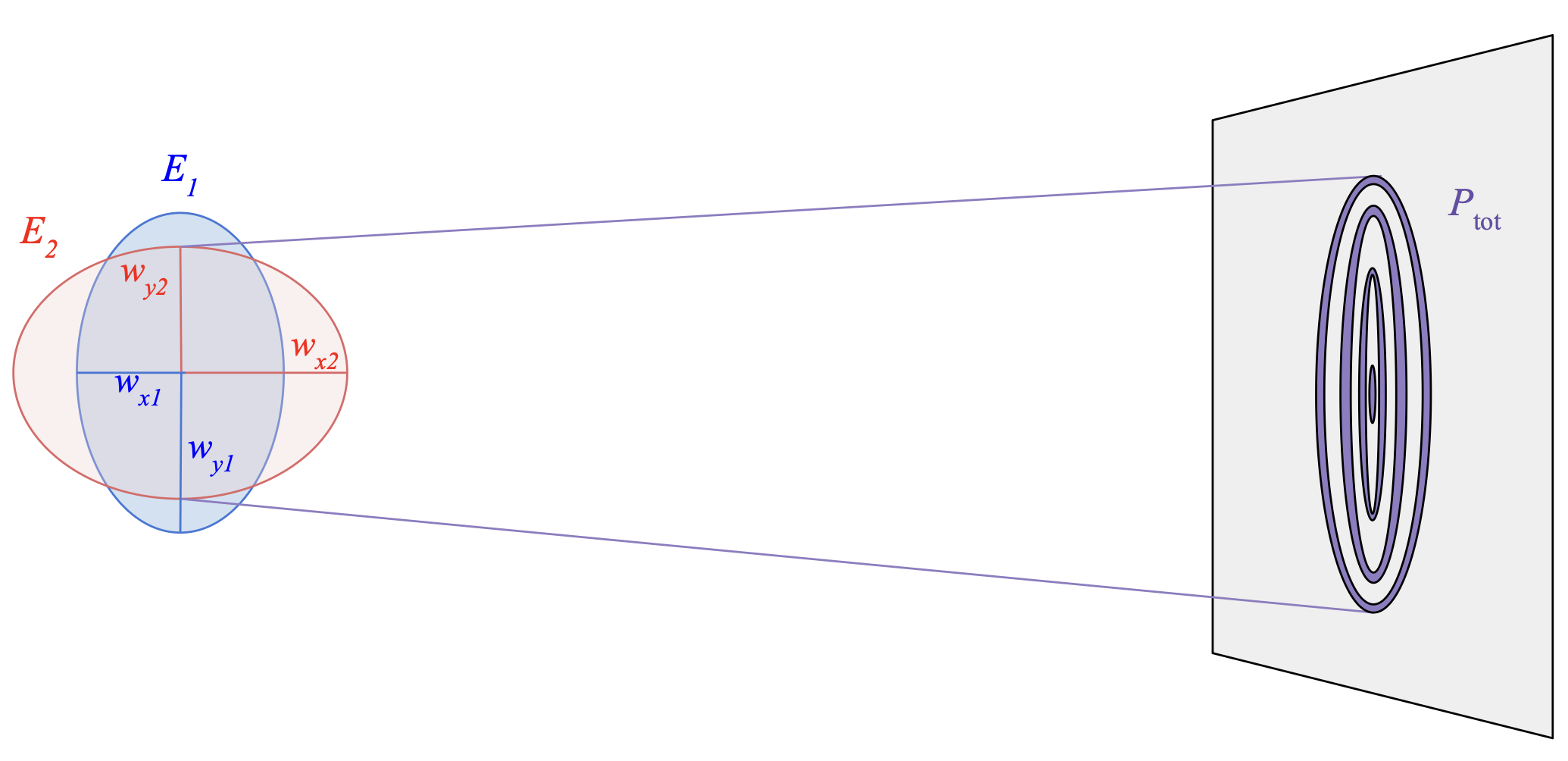}
    \caption{Diagram showing the interference of two elliptical coaxial Gaussian beams. Beam 1 is blue, beam 2 is red, and each is described by two characteristic beam waist parameters (one for each transverse direction). The power of the total beam is shown in purple on a screen, whose cross section should be composed of nested circular fringes.}
    \label{fig:diagram}
\end{figure}

We define \( C \) to be the contrast between the beams, and bound it between 0 and 1. If \( C =1\), then the total power has an amplitude of $P_0$, which means that the beams perfectly interfere. This setup is shown in Fig \ref{fig:diagram}. \newline

If given two elliptical Gaussian fields \( \vec{E}_1 \) and \( \vec{E}_2 \), each with their own beam parameters (power, two beam widths, and two radii of curvature) but approximately equal wavenumbers \( k \), we can write their contrast \( C \) as an analytical function of the ten total beam parameters. This expression is shown in \eqref{result}, and we will derive it from first principles.\newline

We will also test this result by obtaining the beam parameters of a real Michelson interferometer, then comparing its theoretical and experimental values for maximum contrast. 

\section{Derivation}

Suppose two elliptical Gaussian fields \( \vec{E}_1 \) and \( \vec{E}_2 \), each with:
\begin{align*}
    I_j &= |\vec{E}_j(x,y,z)|^2 \\
    P_j &= \int_{-\infty}^{\infty}\int_{-\infty}^{\infty} I_j \ \text{d}x\,\text{d}y = E_{0_j}^2 \frac{w_{0_{x_j}}w_{0_{y_j}}}{2}\pi\ , \quad \text{for} \quad j = 1, 2
\end{align*}

The total field is:
\begin{align*}
    &\vec{E}_{\text{tot}}(x,y,z) = \vec{E}_1 + \vec{E}_2 \\
    &I_{\text{tot}} = |\vec{E}_1 + \vec{E}_2|^2 = I_1 + I_2 + 2\text{Re}[E_1^*E_2]
\end{align*}

So the total power becomes:
\begin{align*}
    P_{\text{tot}} &= \iint \left(I_1 + I_2 + 2\text{Re}\left[A_{xy_1} A_{xy_2}e^{i(\phi_1 - \phi_2)}\right]\right) \text{d}x\,\text{d}y
\end{align*}

\begin{align*}
    A_{xy_1} A_{xy_2} &= E_{0_1}E_{0_2}
    \sqrt{\frac{w_{0_{x_1}} w_{0_{y_1}} w_{0_{x_2}} w_{0_{y_2}}}{w_{x_1} w_{y_1} w_{x_2} w_{y_2}}} \\
    &\quad \times \exp\left[ -x^2\left(\frac{1}{w_{x_1}^2} + \frac{1}{w_{x_2}^2}\right)
    - y^2\left(\frac{1}{w_{y_1}^2} + \frac{1}{w_{y_2}^2}\right) \right] \\ \\
    \phi_1 - \phi_2 &= \frac{k}{2}\Biggl[x^2\left(\frac{1}{R_{x_2}} - \frac{1}{R_{x_1}}\right)
    + y^2\left(\frac{1}{R_{y_2}} - \frac{1}{R_{y_1}}\right) \\
    &\quad - \psi_{x_2} - \psi_{y_2}  + \psi_{x_1} + \psi_{y_1} \biggr]
\end{align*}

Then, if we define the following variables:
\begin{align*}
    P_{12} &\equiv P_1 + P_2 \\
    \psi_{12} &\equiv \frac{1}{2}\left(\psi_{x_1} + \psi_{y_1} - \psi_{x_2} - \psi_{y_2}\right) \\
    a_{12} &\equiv E_{0_1}E_{0_2}
    \sqrt{\frac{w_{0_{x_1}} w_{0_{y_1}} w_{0_{x_2}} w_{0_{y_2}}}{w_{x_1} w_{y_1} w_{x_2} w_{y_2}}} \\
    \alpha_u &\equiv \frac{1}{w_{u_1}^2} + \frac{1}{w_{u_2}^2} + \frac{ik}{2}\left( \frac{1}{R_{u_1}}  - \frac{1}{R_{u_2}} \right), \quad u \in \{x,y\}
\end{align*}

We can write the total power in terms of two Fresnel integrals:
\begin{align}
    P_{\text{tot}} &= P_{12} + 2\text{Re}\left[ e^{i\psi_{12}}\cdot a_{12}
    \int e^{-\alpha_x x^2}\text{dx} 
    \int e^{-\alpha_y y^2}\text{dy} \right]
\end{align}

From the Fresnel identity:
\begin{equation}
    \int_{-\infty}^{\infty} e^{-\alpha x^2} \text{dx} = \sqrt{\frac{\pi}{\alpha}} \quad \text{(valid for complex } \alpha \text{ if Re[}\alpha\text{]}\geq0)
\end{equation}

We find:
\begin{align}\label{complex}
    P_{\text{tot}} = P_{12} + 2\pi a_{12} \ \text{Re}\left[ e^{i\psi_{12}} \cdot \sqrt{\frac{1}{\alpha_x \alpha_y}} \ \right]
\end{align}

Substituting the polar form of complex $\alpha$'s into Equation~\eqref{complex}:
\begin{align*}
    \alpha_u &= |\alpha_u|e^{i\theta_u} \\
    \sqrt{\frac{1}{\alpha_x \alpha_y}} &= \frac{1}{\sqrt{|\alpha_x||\alpha_y|}}e^{-\frac{i}{2}(\theta_x + \theta_y)} \\
    P_{\text{tot}} &= P_{12} + 2\pi \frac{a_{12}}{\sqrt{|\alpha_x||\alpha_y|}} \cos\left(\psi_{12} - \frac{\theta_x + \theta_y}{2} \right)
\end{align*}

\begin{equation}
    \boxed{
    P_{\text{tot}} = P_{12}\left(1 + 2\pi \frac{a_{12}}{P_{12}\sqrt{|\alpha_x||\alpha_y|}}\cos\left(\psi_{12} - \frac{\theta_x + \theta_y}{2} \right)\right)
    }
\end{equation}

Matching this to Equation~\eqref{contrast}, we identify:
\begin{align*}
    C = 2\pi \frac{a_{12}}{P_{12}\sqrt{|\alpha_x||\alpha_y|}}
\end{align*}

Using:
\begin{align*}
    |\alpha_u| &= \sqrt{\left( \frac{1}{w_{u_1}^2} + \frac{1}{w_{u_2}^2} \right)^2 + \frac{k^2}{4}\left( \frac{1}{R_{u_1}} - \frac{1}{R_{u_2}} \right)^2} \\
    a_{12} &= \frac{2}{\pi}\sqrt{\frac{P_1P_2}{w_{x_1} w_{y_1} w_{x_2} w_{y_2}}}
\end{align*}

We finally express the contrast:

\begin{equation}
\boxed{
\begin{aligned}\label{result}
    C &= \frac{4\sqrt{P_1P_2}}{(P_1 + P_2)\sqrt{w_{x_1} w_{y_1} w_{x_2} w_{y_2}}} \\
    &\quad \times \left(\left[\left(\frac{1}{w_{x_1}^2} + \frac{1}{w_{x_2}^2} \right)^2 + \frac{k^2}{4}\left( \frac{1}{R_{x_1}} - \frac{1}{R_{x_2}}\right)^2\right]\right. \\
    &\quad \left.\times \left[\left(\frac{1}{w_{y_1}^2} + \frac{1}{w_{y_2}^2} \right)^2 + \frac{k^2}{4}\left( \frac{1}{R_{y_1}} - \frac{1}{R_{y_2}}\right)^2\right] \right)^{-1/4}
\end{aligned}
}
\end{equation}

In the case where the two beams are identical, sharing the same parameters so that...
\begin{align*}
    P_1 = P_2 &= P \\
    w_{x_1} = w_{x_2} &= w_x \\
    w_{y_1} = w_{y_2} &= w_y \\
    R_{x_1} = R_{x_2} &= R_x \\
    R_{y_1} = R_{y_2} &= R_y
\end{align*}

... we can use Equation~\eqref{result} to show that their contrast is:

\begin{align*}
    C &= \frac{4\sqrt{P^2}}{2P\sqrt{w_x^2w_y^2}}\times\left( \left[\left( \frac{2}{w_x^2} \right)^2 + 0\right] \times \left[\left( \frac{2}{w_y^2} \right)^2 + 0\right] \right)^{-1/4} \\
    &= \frac{2}{w_xw_y}\times \left( \frac{2^4}{w_x^4w_y^4} \right)^{-1/4} \\
    &= 1
\end{align*}

... which is exactly what we expect.

\section{Experimental Methodology}
We now wish to test the experimental validity of~\eqref{result} by computing the maximum theoretical contrast of a real Michelson interferometer and comparing it with its maximum experimental contrast.
\begin{enumerate}
    \item Measure the five elliptical beam parameters of each laser in the interferometer. The beam parameters are:
    \begin{itemize}
        \item Power \(P\)
        \item \(\hat{x}\) beam waist \(w_{0_x}\)
        \item \(\hat{y}\) beam waist \(w_{0_y}\)
        \item \(\hat{x}\) focal position \(z_{0_x}\)
        \item \(\hat{y}\) focal position \(z_{0_y}\)
    \end{itemize}
    
    \item Using the measured beam parameters, the definitions in~\eqref{def}, and expression~\eqref{result}, calculate the maximum theoretical contrast of the interferometer.
    \item Use beam-walking to maximize the experimental contrast of the interferometer and compare it with the theoretical prediction.
    
\end{enumerate}

\subsection{Interferometer setup}
\begin{figure}
    \centering
    \includegraphics[width=\linewidth]{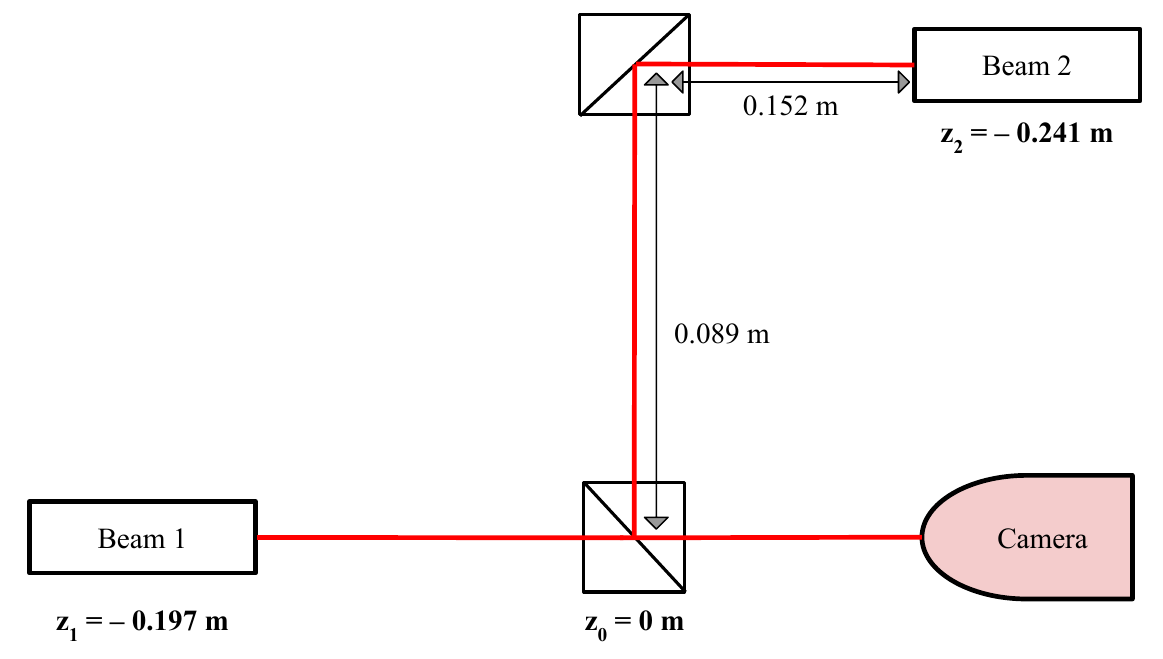}
    \caption{Simple diagram of our Michelson interferometer, where red lines indicate laser paths and squares with a diagonal line indicate beamsplitters. The positions of the beam collimators \(z_{1,2}\) are given with respect to the combining beamsplitter which we defined as \(z_0=0 \text{ m}\).}
    \label{fig:interferometer}
\end{figure}

\begin{figure}
    \centering
    \includegraphics[width=\linewidth]{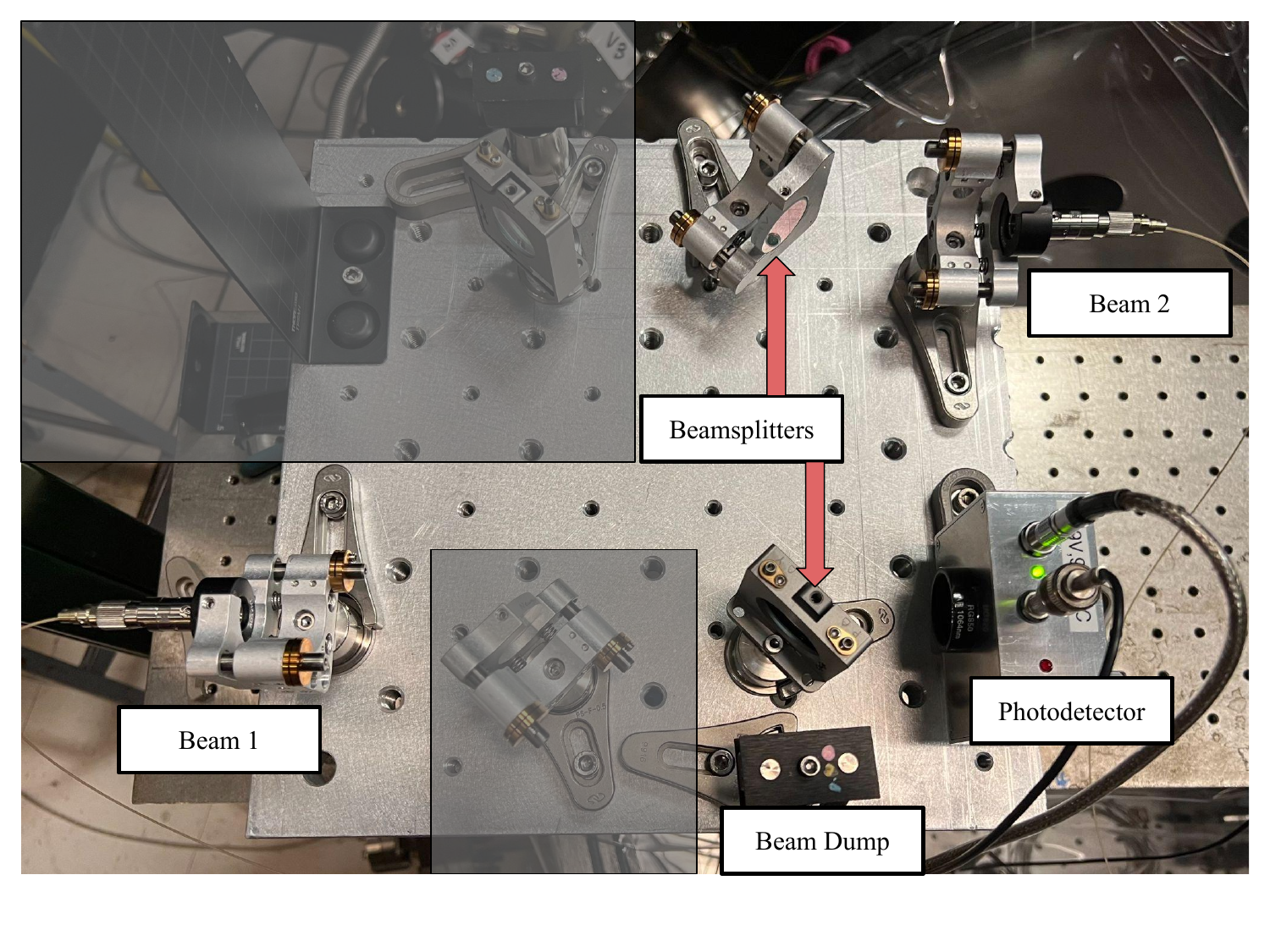}
    \caption{Photograph of the Michelson interferometer used in our experiment. The two beam collimators, two beamsplitters, and photodetector used to determine contrast are shown and labeled. Extraneous beamsplitters and beam dumps not used in this experiment are partially grayed-out.}
    \label{fig:picture}
\end{figure}

Our free-space Michelson interferometer combines two nominally TEM\textsubscript{00} beams at wavelength \(\lambda = 1064 \, \text{nm}\). These beams are additionally modulated by frequencies \(f_1 = 80 \,\text{MHz}\) and \(f_2 = 80 \, \text{MHz} + 100 \, \text{Hz}\) using AeroDiode fiber-coupled acousto-optic modulators, giving an optical beat note of \(100 \, \text{Hz}\). The beams are emitted from Schäfter+Kirchhoff 60FC-0-A18-03-Ti-V fiber-coupled collimators, and are passed through Newport 10B20NP.29 fifty-fifty beamsplitters.\\ \\
We define the recombination plane as \(z_0 = 0\), then beam 1's collimator is at \(z_1 = -0.197 \,\text{m}\) and beam 2's collimator is effectively at \(z_2 = -0.241\,\text{m}\). The powers of each beam were measured independently with a Thorlabs S130C power sensor, while the mode measurements were taken by a WinCamD-LCM beam profiling camera. Fig \ref{fig:interferometer} depicts a basic diagram of the setup, while Fig \ref{fig:picture} shows an image of it.\newline

\subsection{Measuring beam parameters}
For each beam and axis (\(\hat{x}\), \(\hat{y}\)), we used a WinCamD to acquire 1D intensity profiles across five axial planes \(z_i\). Using the definition of \(z_0\) above, the five measurements planes were \(z_i \in [0.051\,\text{m}, 0.159\,\text{m}, 0.324\,\text{m}, 0.533\,\text{m}, 0.787\,\text{m}]\). For each beam at each distance \(z_i\), we take \(\hat{x}\) and \(\hat{y}\) beam profile images using the WinCamD by DataRay. Each image is the average  over 10 frames, with an exposure time of 84.90 milliseconds each.\newline

\begin{figure}
    \centering
    \includegraphics[width=\linewidth]{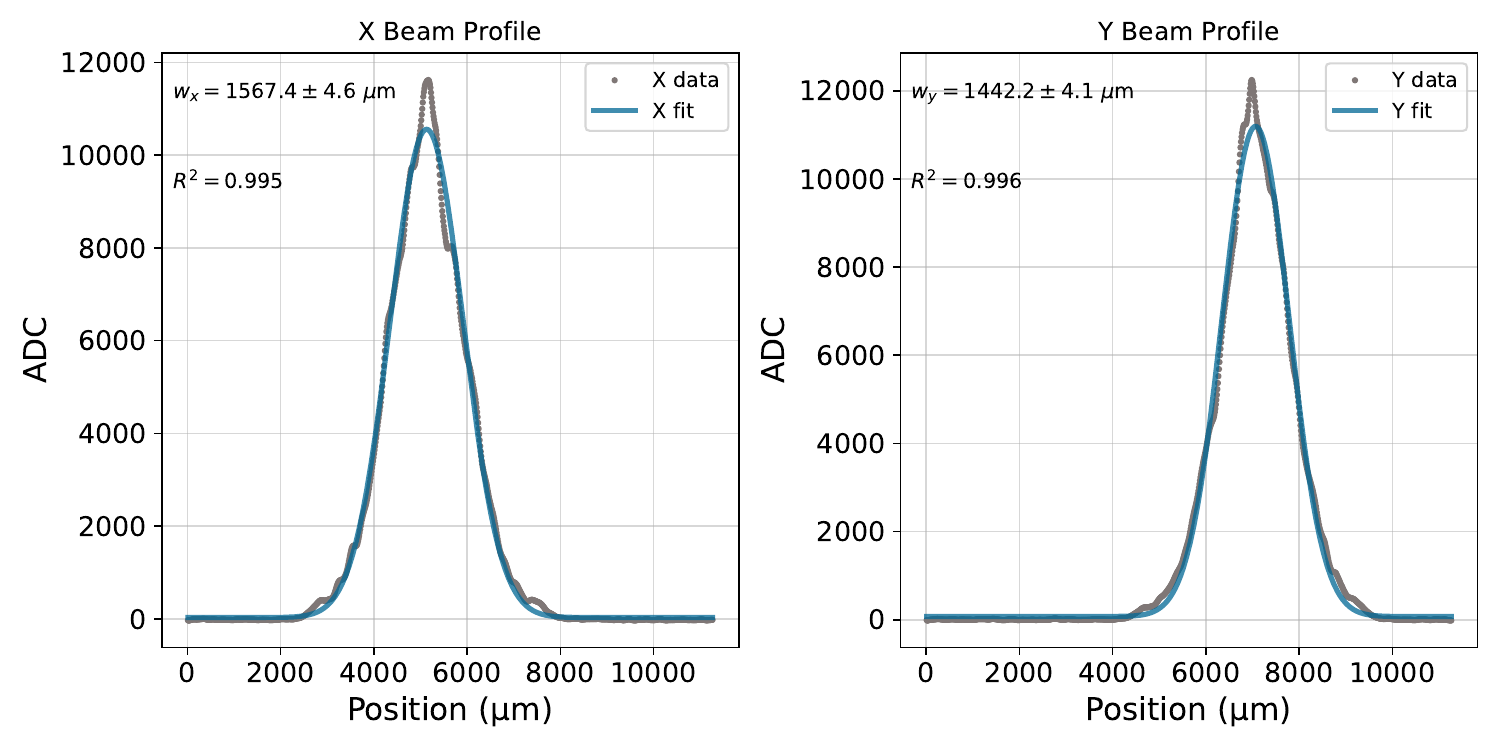}
    \caption{The \(\hat{x}\) and \(\hat{y}\) profiles of beam 1 in our laser interferometer at the position \(z = 0.051 \; \text{m}\). Gray scatter points show the actual ADC values measured by our camera, while the blue curves show the Gaussians-of-best-fit for each profile according to \eqref{eq:gauss}. The beam widths \(w_{x,y}\) of each Gaussian is printed with its uncertainty. The correlation coefficient \(R^2\) between each Gaussian and its raw data is also shown.}
    \label{fig:profile}
\end{figure}

We then use nonlinear weighted least squares to fit each beam profile to a 1D Gaussian of the form:
\begin{align}
    I(u,z_i) = I_0(z_i)\exp\left[-2 \left(\frac{u-u_0(z_i)}{w_u(z_i)}\right)^2\right] + B(z_i) \label{eq:gauss}
\end{align}
where \(u \in \{x,y\}\), \(I_0\) is the intensity of the Gaussian peak, \(u_0\) is the position of the Gaussian peak, \(w_u\) is the radius of the beam where the intensity drops by \(1/e^2\), and \(B\) is a background offset. Fig \ref{fig:profile} illustrates an example of this process. \newline

\begin{figure}
    \centering
    \includegraphics[width=\linewidth]{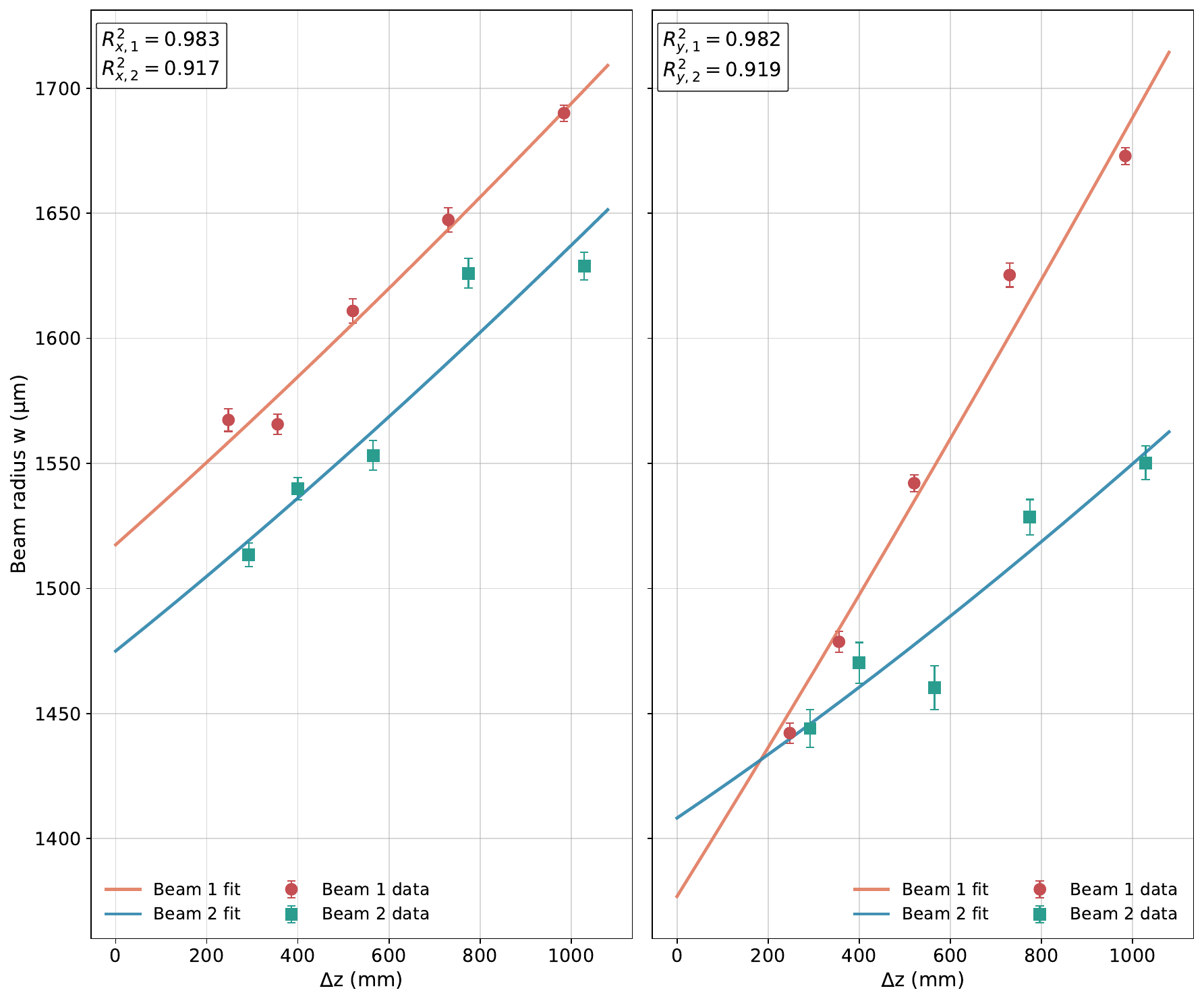}
    \caption{The left panel shows the five \(\hat{x}\) beam width measurements of each beam, and their curve of best fit as according to \eqref{def}. The right panel does the same, but for the \(\hat{y}\) direction instead. The x-axis of each panel shows the distance between the camera and the beam collimator, denoted as focal displacement \(\Delta z\). Finding focal position of the \(j\)th beam from its focal displacement is simply \(z_{0_j} = \Delta z + z_j\) where \(z_{1,2}\) are defined above. The four correlation coefficients for the width function of the \(j\)th beam in the \(\hat{u}\) direction, labeled as \(R^2_{u, j}\), are printed as well. The eccentricities of beams 1 and 2 are 0.694 and 0.123 respectively.}
    \label{fig:fitting}
\end{figure}

Then for each beam profile we have five fitted beam width values with respective uncertainties. Using the definition of beam radius in~\eqref{def} we use weighted least squares once again to fit for beam waist \(w_{0_u}\) and focal position \(z_{0_u}\) as shown in Fig \ref{fig:fitting}. This allows us to calculate  four parameters \(\{w_{0_x}, w_{0_y}, z_{0_x}, z_{0_y}\}\) and their errors for each beam. We also used a power sensor to measure the power of each beam, obtaining the final pair of parameters.

\subsection{Calculating theoretical contrast}

Using the methodology described above, we can obtain the values and uncertainties of all ten beam parameters in our interferometer, which are shown in Table \ref{tab:params}. Now, we can use equations \eqref{def} and \eqref{result} to calculate the maximum theoretical contrast. The value obtained for this interferometer is 0.968 or 96.8\%.\newline

\begin{table}[h!]
    \centering
    \begin{tabular}{lccc}
    \toprule
    Name & Symbol & Value & Units \\
    \midrule
    Beam 1 \(\hat{x}\)-waist & \(w_{0_{x_1}}\) & \(1229.13 \pm 18.59\) & \(\mu\)m  \\
    Beam 1 \(\hat{y}\)-waist & \(w_{0_{y_1}}\) & \(884.83 \pm 12.05\) & \(\mu\)m \\
    Beam 1 \(\hat{x}\)-focal pos. & \(z_{0_{x_1}}\) & \(-3436.7 \pm 22.0\) & mm  \\
    Beam 1 \(\hat{y}\)-focal pos. & \(z_{0_{y_1}}\) & \(-2953.0 \pm 24.6\) & mm \\
    Beam 1 Power & \(P_1\) & \(7.5\pm 0.2\) & mW \\
    \midrule
    Beam 2 \(\hat{x}\)-waist & \(w_{0_{x_2}}\) & \(1244.70 \pm 24.33\) & \(\mu\)m  \\
    Beam 2 \(\hat{y}\)-waist & \(w_{0_{y_2}}\) & \(1254.21 \pm 31.08\) & \(\mu\)m \\
    Beam 2 \(\hat{x}\)-focal pos. & \(z_{0_{x_2}}\) & \(-3149.5 \pm 50.9\) & mm  \\
    Beam 2 \(\hat{y}\)-focal pos. & \(z_{0_{y_2}}\) & \(-2612.4 \pm 103.4\) & mm \\
    Beam 2 Power & \(P_2\) & \(8.4 \pm 0.2\) & mW \\
    \bottomrule
    \end{tabular}
    \caption{The name, mathematical symbol, measured value, and measured uncertainty of all ten beam parameters of the Michelson interferometer used in our experiment.}
    \label{tab:params}
\end{table}

To find the contrast uncertainty, we use the standard formula for the propagation of error. To write this, we first define our parameter vector:

\[
\vec{\theta} = \begin{bmatrix}
w_{0_{x_1}} & w_{0_{y_1}} & z_{0_{x_1}} & z_{0_{y_1}} & w_{0_{x_2}} & w_{0_{y_2}} & z_{0_{x_2}} & z_{0_{y_2}} & P_1 & P_2
\end{bmatrix}^\top \in \mathbb{R}^{10}
\] \\
Let \(C(\vec{\theta})\) define the maximum contrast of the interferometer at \(z=0\). Then we define the Jacobian vector of \(C\) at the mean value \(\vec{\mu} =\mathbb{E}[\vec{\theta}]\) as:

\[
\vec{J} \equiv \vec{\nabla}_{\theta}\, C(\vec{\theta})|_{\vec{\theta}=\vec{\mu}} = \begin{bmatrix}
    \frac{\partial C}{\partial w_{0x1}} & \frac{\partial C}{\partial w_{0y1}} & \cdots & \frac{\partial C}{\partial P_1} & \frac{\partial C}{\partial P_2} 
\end{bmatrix}^\top \in \mathbb{R}^{10}
\] \\
We also define the covariance matrix \(S\) as follows:
\[
S \equiv \text{Cov}(\vec{\theta}) \in \mathbb{R}^{10\times10}\, , \quad S_{ij} = \text{Cov}(\theta_i, \theta_j)
\] \\
Then to first order, the variance and uncertainty for contrast are \cite{Taylor1997}:
\begin{align}
\text{Var}(C) = \vec{J}\cdot S\vec{J} \, , \quad \sigma_{C, \text{theory}} = \sqrt{\vec{J}\cdot S\vec{J}} \label{eq:err}
\end{align} \\
Using this process, we found that \(\sigma_{C, \text{theory}} = 0.005\). Thus the theoretical maximum contrast of our interferometer, with an uncertainty of 1\(\sigma\), is:
\begin{align}
    \boxed{C_{\text{theory}} = 0.968 \pm 0.005} \label{eq:theory}
\end{align}

\subsection{Finding experimental contrast}
\begin{figure}
    \centering
    \includegraphics[width=\linewidth]{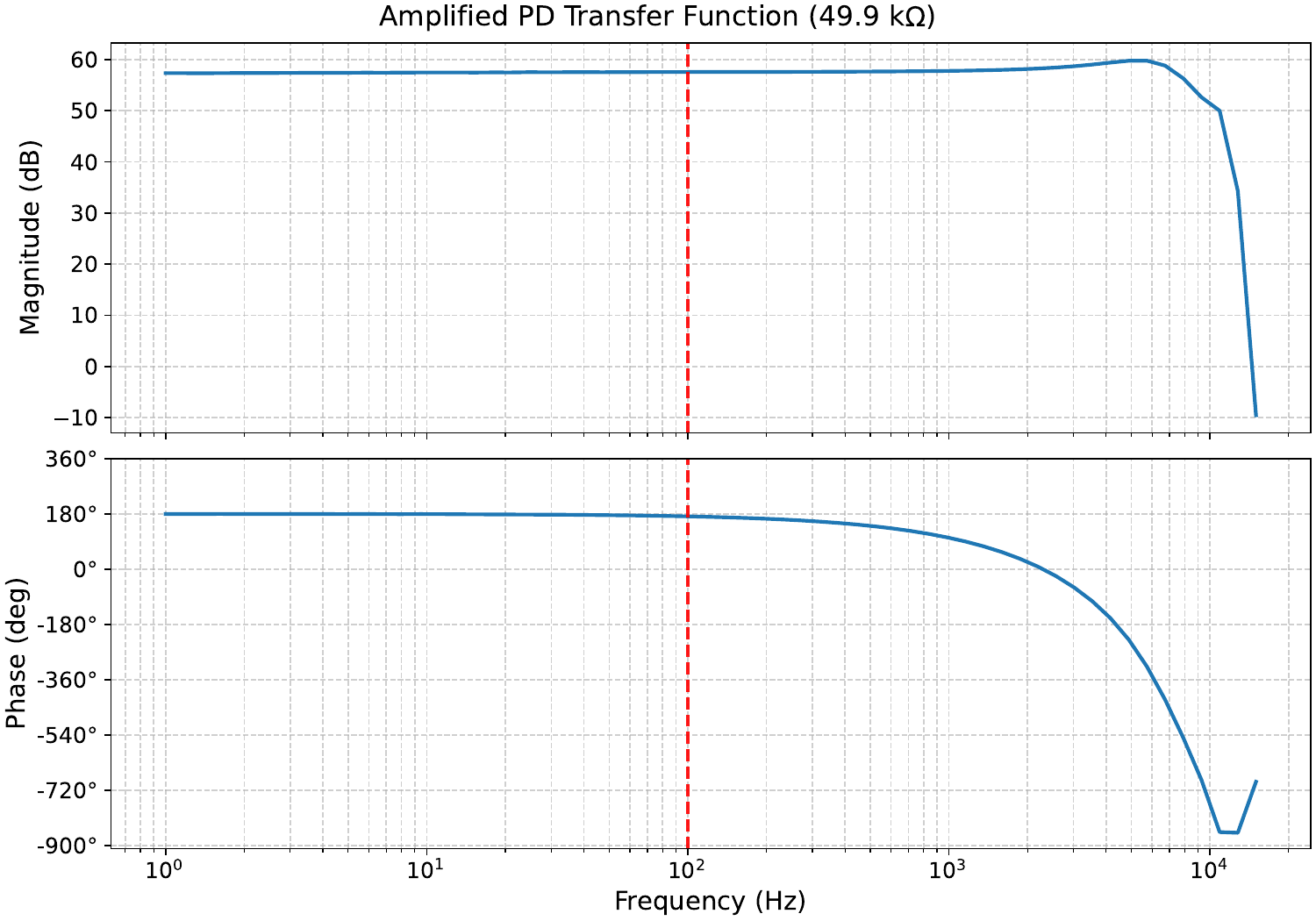}
    \caption{Bode plots of our amplified photodetector's transfer function. The heterodyne frequency of 100 Hz lies within the range where the system response is flat and close to unity. The amplitude plot has arbitrary units of magnitude, and the photodetector's transimpedance amplifier was set to \(49.9 \text{ k}\Omega\).}
    \label{fig:transfer}
\end{figure}

To maximize the actual experimental contrast of our interferometer, we performed beam-walking by tuning the pitch and yaw of the two beamsplitters shown in Fig \ref{fig:interferometer}. We replaced the WinCamD-LCM with a transimpedance amplified photodetector connected to an Tektronix TDS 2024B oscilloscope, to measure the minimum and maximum voltages of the signal. We then measured the transfer function between the modulation input voltage of the test laser and the output voltage of our amplified photodetector as measured by an oscilloscope. The transfer function is shown in Fig \ref{fig:transfer}, with the heterodyne frequency of 100 Hz marked in red.

We use an experimental formula for contrast \cite{Paschotta_2005_interference}:
\begin{align}
    C = \frac{V_{\text{max}} - V_{\text{min}}}{V_{\text{max}} + V_{\text{min}}} = \frac{V_{\text{peak-peak}}}{2V_{\text{mean}}} \label{eq:exp_contrast}
\end{align}
With no signal at all, the oscilloscope's zero-current reading has a peak-to-peak voltage of 30 mV. Then the uncertainty of \(V_{\text{max,min}}\) is \(\sigma = 0.015 \text{ V}\). Using propagation of error:
\begin{align*}
    &\sigma_{\text{peak-peak}} = \sigma\sqrt{2} = 0.021 \text{ V} \\
    &\sigma_{\text{mean}} = \frac{\sigma}{\sqrt{2}} = 0.011 \text{ V}  \\
    &\sigma_{C,\text{exp}} = \frac{\sigma}{2V_{\text{mean}}}\sqrt{2 + \frac{1}{2}\left(\frac{V_{\text{peak-peak}}}{V_\text{mean}}\right)^2} = 0.007 \text{ V}
\end{align*}
After beam-walking to find the highest peak-to-peak reading:
\begin{align*}
    V_{\text{peak-peak}} = 4.20\pm 0.021\text{ V}\\
    V_{\text{mean}} = 2.21\pm 0.011\text{ V}
\end{align*}
Thus, the maximum experimental contrast of our interferometer is:
\begin{align}
    \boxed{C_\text{exp} = 0.950 \pm 0.007} \label{eq:exp}
\end{align}
Then the mean theoretical prediction in \eqref{eq:theory} overestimates the mean experimental result by 1.89\%.

\section{Discussion}
Because \(C_\text{theory}\) and \(C_\text{exp}\) lie outside of each others' confidence intervals, it is worth examining additional sources of experimental error not considered in our theory. We believe the most important ones to be:
\begin{itemize}
    \item Our formula \eqref{result} assumes that the elliptical cross sections of the two interfering beams share the same center and axes. If one of the physical beams is rotated with respect to the other, or translated so that they do not share a center, the theoretical prediction \eqref{eq:theory} will be an overestimate.
    \item The distances at which each mode measurement was taken were measured with a meter stick with 1/32 inch tick marks. Thus, each measurement in Fig \ref{fig:fitting} should also have horizontal error bars on the order of \(\sim1\) mm. This introduces additional covariance to the calculation of \(\sigma_C\) and will increase the uncertainty range. 
    \item As seen in Fig \ref{fig:fitting}, the propagation of beam 2 does not adhere closely to its curve of best fit, implying that there might be aberrations from higher order modes or other optical effects. These factors would likely cause the interference of beams 1 and 2 to deviate from the expected pattern, thus causing our theory to provide an overestimate.
    \item As seen in Fig \ref{fig:profile}, the real beam profiles are not entirely Gaussian. In both the \(\hat{x}\) and \(\hat{y}\) profiles, there exist sharp ridges which introduce interference patterns that our theory does not account for. By ``smoothing" these ridges out, \eqref{eq:theory} is likely an overestimate.
    \item Each pixel's ADC reading in Fig \ref{fig:profile} is averaged over ten frames, meaning every data point has a standard error that hasn't been accounted for. This is another source of covariance that would increase the uncertainty range in \eqref{eq:theory}.
\end{itemize}
Because our experimental error was only 1.89\% despite these additional factors, we believe that our analytical formula in equation \eqref{result} is an effective theoretical predictor of the contrast for an interferometer. However, from our testing we also recognize the limitations of its assumptions, and demonstrate why it might overestimate the true experimental contrast.

\section{Acknowledgments}
The authors gratefully thank the National Science Foundation for their support. LIGO was constructed by the California Institute
of Technology and the Massachusetts Institute of Technology with funding from the National Science Foundation and operates
under cooperative agreement PHY-0757058. Advanced LIGO was built under award PHY-0823459. This document has been assigned LIGO Laboratory document
number LIGO-P2500734. Public internal LIGO documents are found at: https://dcc.ligo.org/cgi-
bin/DocDB/DocumentDatabase/.

\printbibliography

\end{document}